# An Extension Of Weiler-Atherton Algorithm To Cope With The Self-intersecting Polygon

**ABSTRACT:** In this paper a new algorithm has been proposed which can fix the problem of Weiler-Atherton algorithm. The problem of Weiler-Atherton algorithm lies in clipping self-intersecting polygon. Clipping self-intersecting polygon is not considered in Weiler-Atherton algorithm and hence it is also a main disadvantage of this algorithm. In our new algorithm a self-intersecting polygon has been divided into non-self-intersecting contours and then perform the Weiler-Atherton clipping algorithm on those sub polygons. For holes we have to store the edges that is not hole contour's own boundary from recently clipped polygon. Thus if both contour is hole then we have to store all the edges of the recently clipped polygon. Finally the resultant polygon has been produced by eliminating all the stored edges.

**INTRODUCTION :** Clipping 2D polygons is one of the important task in the computer graphics. In case of 3D polygon rendering this task has to be done several times. In modern applications more general and complex polygon has to be clipped. But only a few algorithms too general to clip such polygon. Sutherland and Hodgeman's algorithm[2] is limited to convex clip polygons. The algorithm presented in [6] i.e Liang-Barsky are able to clip polygon but the problem lies on the clipper polygon's limitation i.e clipper polygon must have to be rectangle. More general algorithms were presented in [8, 9, 10,11]. They allow concave polygons with holes, but they do not permit self-intersections. Vatti [5] and Greiner-Horman [1] algorithm can able to solve this problem that means these algorithms are general enough to be able to clip any arbitrary polygons. But with this algorithm Weiler-Atherton algorithm [3] can also be able to clip self-intersecting polygons.

Greiner-Horman [1] algorithm uses less complex data structure and also able to clip polygon in reasonable time. But there is a limitation in Greiner-Horman [1] algorithm. As in Greiner-Horman [1] algorithm, hole contour of a self-intersecting polygon can only be calculated through even-odd winding number, it unable to produce right result in case of a self-intersecting polygon which is calculated by non-zero winding rule. Again Greiner-Horman [1] algorithm suffers from degeneration problem though this problem can be solved by the algorithm presented in [7] but then space complexities increases significantly. Weiler-Atherton does not suffer from degeneration problem.

**PROPOSED METHODOLOGY :** This algorithm is based on winding number of a contour. From the winding number we have to calculate the hole information of a self-intersecting polygon.

## Winding Number :-
The winding number of a polygon (contour) C about a point x,w measures not only whether C encloses x, but also how many times and in which orientation C "winds around" x.

In particular

$$W = \begin{cases} 0 & \text{if x is not inside C} \\ n > 0 & \text{if C winds around x in n times counter clockwise} \\ n < 0 & \text{if C winds around x (- n) times clockwise} \end{cases}$$

The winding number can be rigorously defined as a contour integral in the complex plane [12].

$$w = \frac{1}{2\pi i} \int_C \frac{1}{z} dz \quad \text{where } z = x+iy$$

Winding number is not defined when the point x is on the polygon C.

## Axis Crossing Method :-

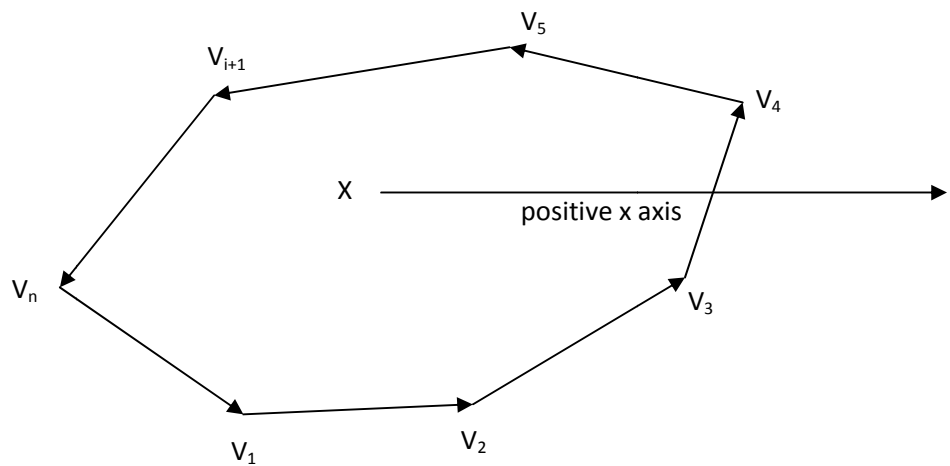

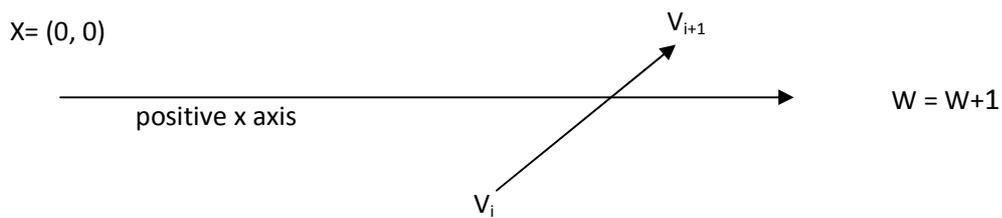

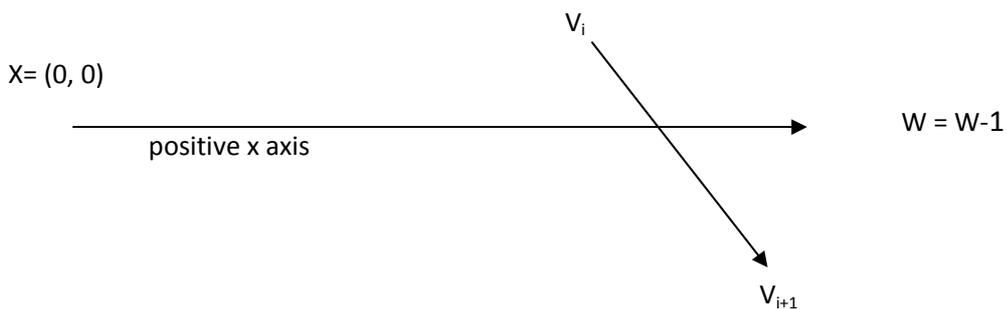

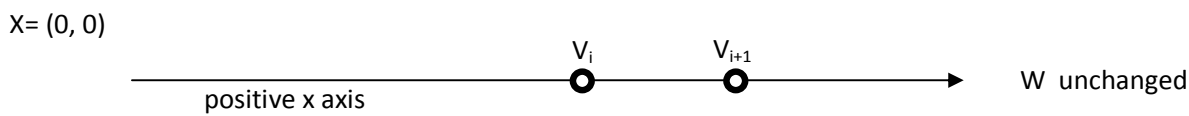

There is two types of winding number calculation rule. First is even-odd winding rule where contour having even winding number is considered as hole and contour having odd winding number is considered as sub-contour. Second non-zero winding rule where contour having only zero winding number is considered as hole else it is considered as sub-contour.

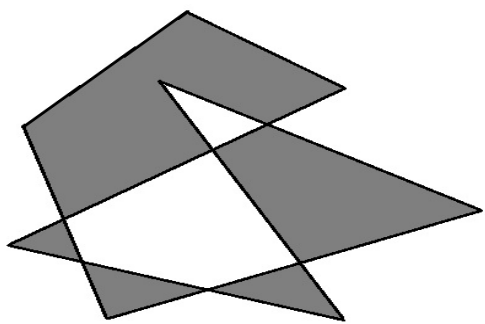
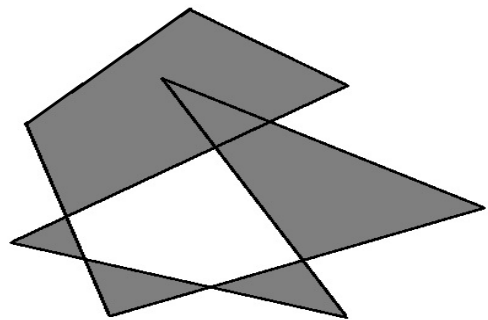

Fig (a)                                   Fig (b)

The interior of a self-intersecting polygon based on even-odd rule [Fig (a)] and non-zero winding number [Fig (b)].

**Problem in Weiler-Atherton Algorithm :** To evaluate this algorithm first we have to realize the problem in Weiler-Atherton algorithm.

Let take an example for better understand.

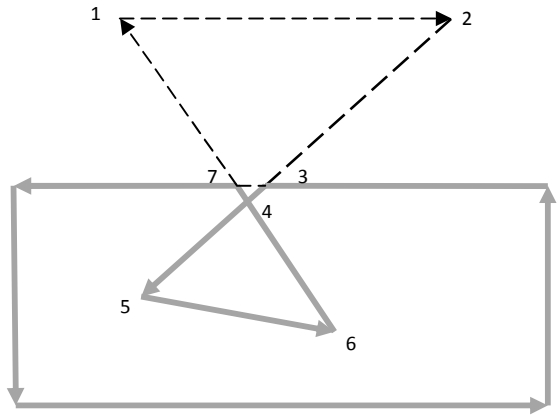

Incorrect output in Weiler-Atherton algorithm

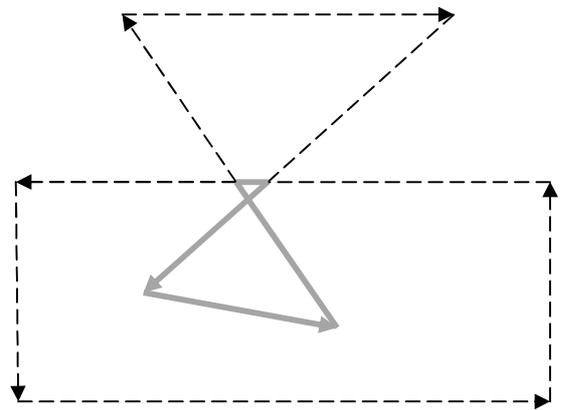

Correct

Starting from 1 we enter into the clipper (i.e the rectangular polygon) polygon at 3 and start printing edges 3-4, 4-5, 5-6, 6-4, 4-7. At 7 as it is leaving the polygon it will follow the edge of the clipper polygon & start printing edges of the clipper polygon. Hence it produce wrong output.

The actual problem is that we can't determine the exact clock orientation of a self-intersecting polygon. And as Weiler-Atherton algorithm is very much dependent of clock-orientation it produces erroneous output in case of self-intersecting polygon.

**THE ALGORITHM :** In this algorithm we have to subdivide the self intersecting polygon into some non-self-intersecting sub polygons. We have to do this for both polygon i.e subject polygon and clipper polygon if both polygons are self-intersecting.

So first we have to consider the division of a self-intersecting polygon. We can use stack or linklist for this. Now when we traverse a vertex we store those vertexes into the stack unless a match occurs (i.e the same vertex is already in the array). If match occurs then pop up the vertex and store it in another array or linklist as a contour until we get the matched vertex.

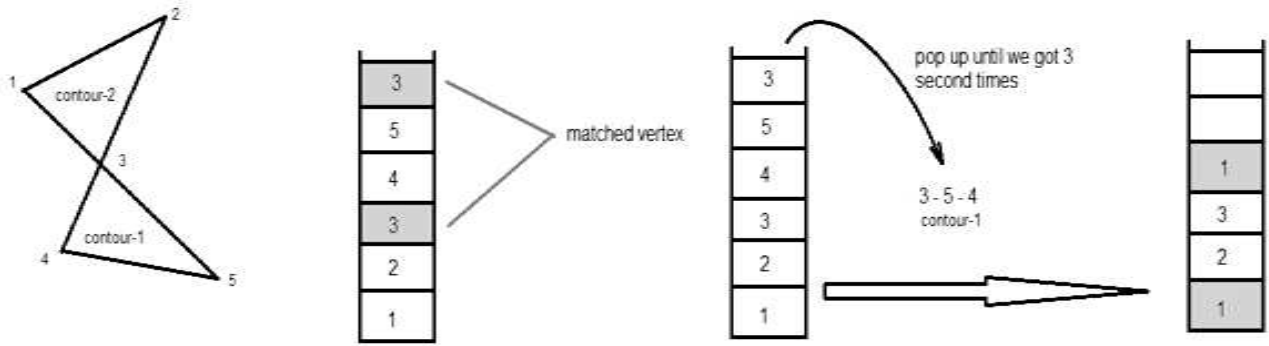

Thus we get a contour and continue this step until all vertexes are traversed.

Main problem of clipping self-intersecting polygon using Weiler-Atherton algorithm is the clock orientation. It can be easily understand from the picture. For contour-1 the clock orientation is counter-clockwise but for contour-2 the clock orientation is clockwise. Now as Weiler-Atherton algorithm is strictly dependent on clock orientation we have to fix clock orientation of contours.

According to Weiler-Atherton algorithm the clipper polygon and subject polygon must have the same clock orientation and hole must have opposite clock orientation.

To determine the clock orientation we can use winding number calculation –

Winding number > 0         its anti-clockwise.

Winding number < 0         its clockwise

So in case of all contours of both polygons which is not hole should have same clock orientation and for the all hole contour of polygon should have opposite clock orientation.

If a contour is in opposite clock orientation of what it should be we can change its clock orientation by simply reversing the array or linklist of the edge list.

We have to obtain hole information for both polygons. This can be easily done by winding numbers rule. If winding number of a contour is 0 then the contour is considered as hole. The hole contours have to be stored and will be used in time of clipping.

Now as we get the contours we can clip the sub-polygons of the subject polygon with the sub-polygons of the clipper polygon. And resultant polygon will be stored temporarily.

So for any polygon the general algorithm will be –

**1:** for (each contour of the subject polygon)

**2:**     If the contour is not hole.

**3:**         for (each contour of the clipper polygon)

**4:**             If the contour is not hole.

**5:**                 perform Weiler-Atherton algorithm with considering the hole contours.

**6:**                 delete those edges from recently clipped polygon which fall inside the hole contours and print the remaining edges (this can be done if mid-point of an edge falls inside a hole).

**7:**             End if

**8:**         End for

**9:**     End if

**10:** End for

## experimental result :

In this algorithm for the following example corresponding steps are presented.

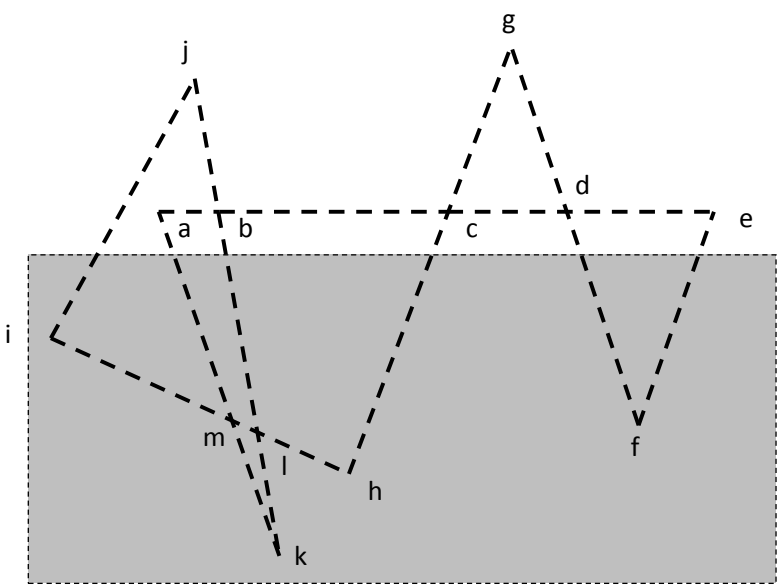

Here a-b-l-m-a contour is calculated by non-zero winding rule. Hence it will not be considered as hole.

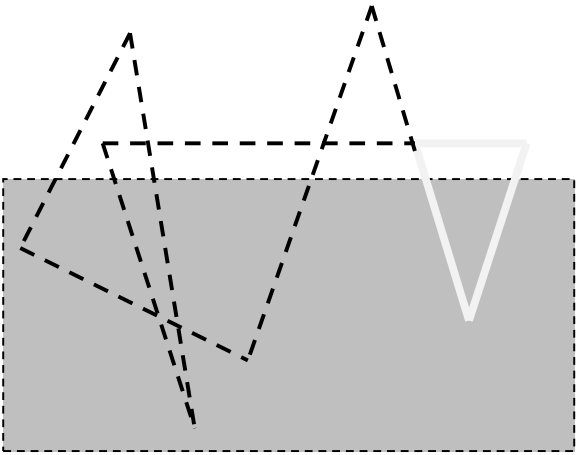 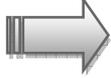 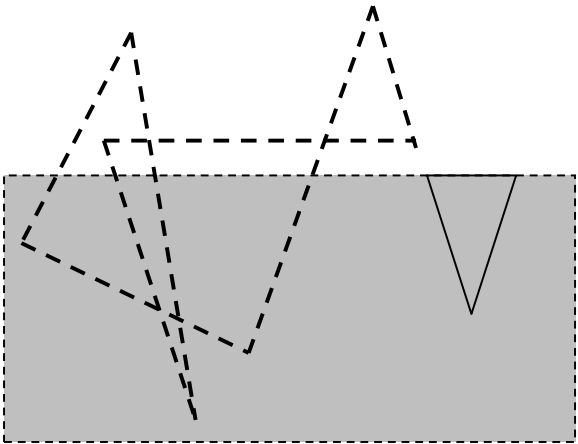
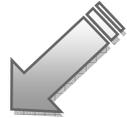
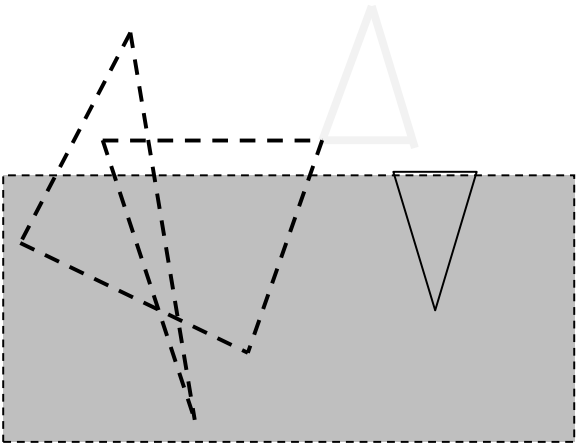 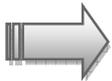 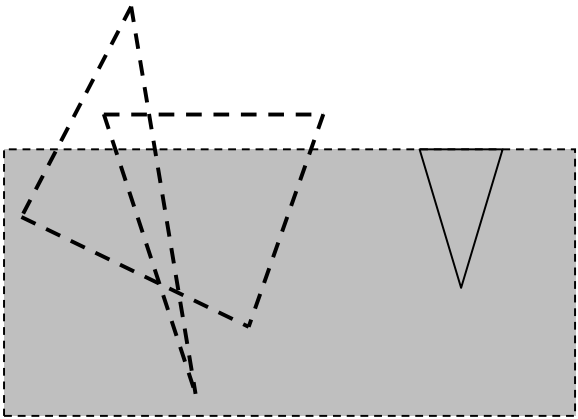
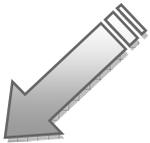
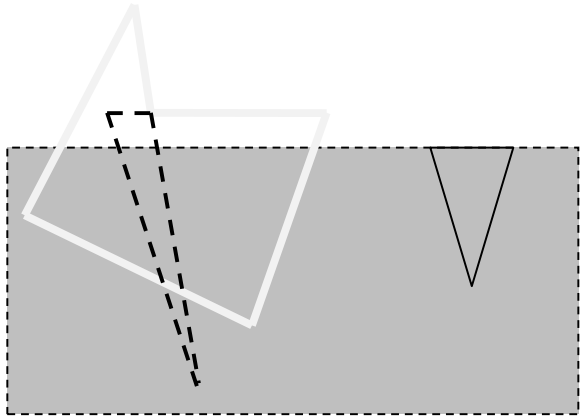 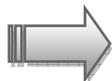 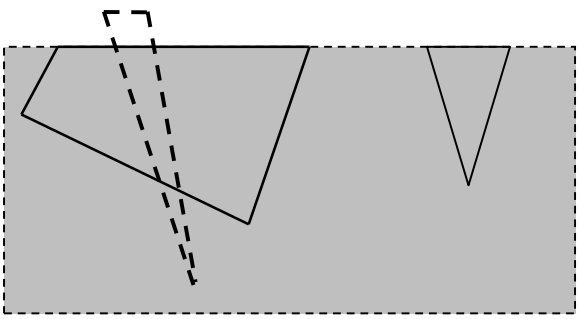
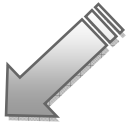
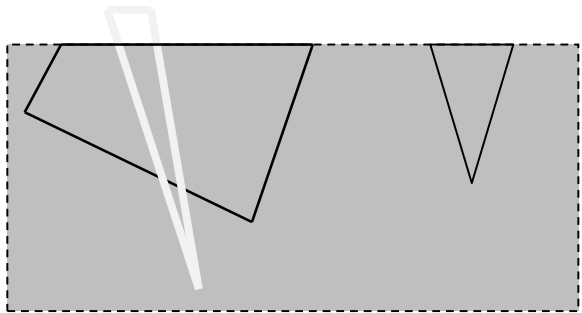 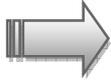 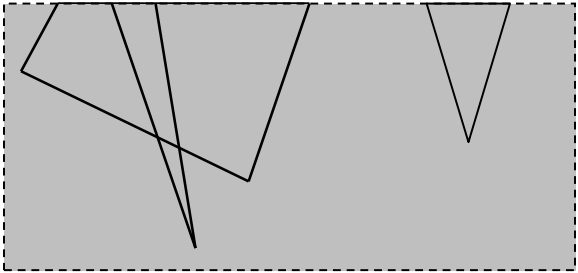

**Conclusion :** Though Greiner-Horman algorithm [1] works fine but if the hole is calculated by non-zero winding rule it can not clip the the self-intersecting polygon properly. But through this algorithm we can clip self-intersecting polygon with hole calculated by both even-odd winding rule and non-zero winding rule. ***But most importantly through this algorithm not only Weiler-Atherton algorithm but also any algorithm that can clip a general concave polygon will be able to clip any self-intersecting polygon.***

**Acknowledgement :** I wish to thank Dr. Shantanu Halder for his kind co-operation in this project without his help this project never would have succeeded.